\documentclass{elsart}
\usepackage{natbib,epsf,rotate,subfigure}

\begin{document}
\runauthor{Kirchner}

\begin{frontmatter}
\title{Optimization of drift gases for accuracy in pressurized drift tubes}
\author{J. J. Kirchner,}
\author{U. J. Becker,} 
\author{R. B. Dinner,} 
\author{K. J. Fidkowski,}
\author{and J. H. Wyatt}

\address{Massachusetts Institute of Technology \\ Laboratory of Nuclear Science \\ Cambridge, MA 02139 USA}

\begin{abstract}
Modern detectors such as ATLAS use pressurized drift tubes to minimize diffusion and achieve high coordinate accuracy.  However, the coordinate accuracy depends on the exact knowledge of converting measured times into coordinates.  Linear space-time relationships are best for reconstruction, but difficult to achieve in the $E \propto \frac{1}{r}$ field.  Previous mixtures, which contained methane or other organic quenchers, are disfavored because of ageing problems.  From our studies of nitrogen and carbon dioxide, two mixtures with only small deviations from linearity were determined and measured.  Scaling laws for different pressures and magnetic fields are also given.
\end{abstract}
\end{frontmatter}

\section{Introduction}
Since the invention of Geiger counters, the simplicity of proportional tubes has been attractive.  However, the inhomogeneous electric field $E \propto \frac{1}{r}$ poses a problem for accurate spatial measurements because the drift velocities depend on electric field \cite{bib:sysstudy}.  This means a complicated relation between the measured time and the radius inferred for reconstruction.  A linear relation with small correction provides accurate tracking and was previously achieved by mixtures with methane \cite{bib:atlas_tdr}.

The search for an optimized drift gas will be discussed in terms of the ATLAS Monitored Drift Tube (MDT) system, which uses 370 000 drift tubes in 1 194 chambers with an expected spatial resolution of 80 $\mu$m \cite{bib:atlas_tdr}.  While the originally proposed gas mixture Ar:CH$_4$:N$_2$ (91:5:4)\% met this resolution requirement, it was found that mixtures with hydrocarbon admixtures such as methane are prone to ageing \cite{bib:herten,bib:ageing,bib:riegler}.  The mixture Ar:CO$_2$ (93:7)\% was subsequently used because of its resistance to ageing.

To achieve high accuracy in coordinate reconstruction, it is important to keep corrections small.  This equates to having a linear space-time relation, so that the electron mean drift velocity is nearly independent of electric field.  Also, in large systems operating for long periods of time, the drift gas must have high tolerance for cumulative contamination by nitrogen from air, whereas oxygen is usually removed by purifiers.  Surprisingly, the effects of small amounts of nitrogen are significant \cite{bib:n2_cont}, but can be used advantageously by using nitrogen as an intentional admixture in the gas mixture \cite{bib:inelastic}.

\begin{figure}[ht]
\centerline{
\epsfxsize=3.5in
\epsfbox{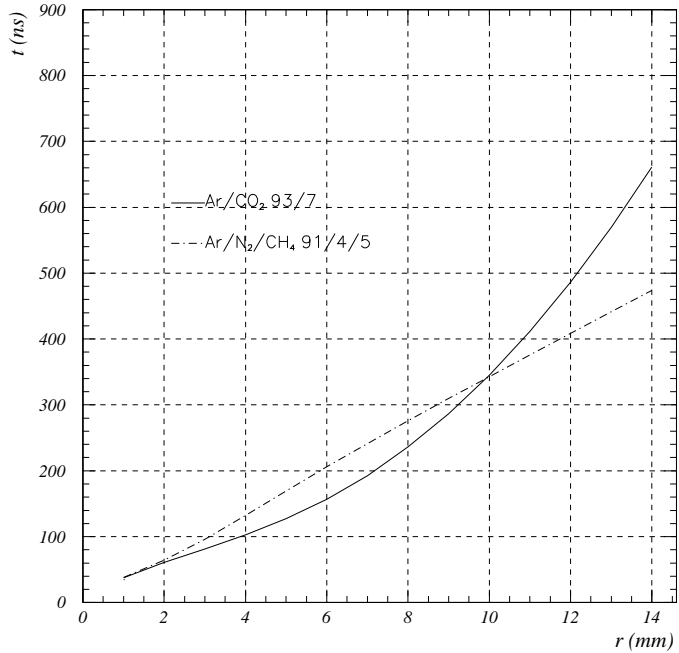}}
\caption{Space-time relations for different gas mixtures \cite{bib:riegler}.  The original design gas was Ar:CH$_4$:N$_2$ (91:5:4)\% and the present operational gas is Ar:CO$_2$ (93:7)\%.} \label{fig:space-time}
\end{figure}

As seen in Fig.~\ref{fig:space-time}, the space-time relation for the original ATLAS mixture Ar:CH$_4$:N$_2$ (91:5:4)\% is very linear unlike the relation for the current test mixture of Ar:CO$_2$ (93:7)\%.  

Knowledge of properties of electron drift in gaseous detectors \cite{bib:cyclo} will allow for a systematic search for a gas mixture which improves coordinate accuracy.  To avoid ageing, it is important to find a mixture with a linear space-time relation which does not include a hydrocarbon mixture.

\section{Experimental setup}
\begin{figure}[ht]
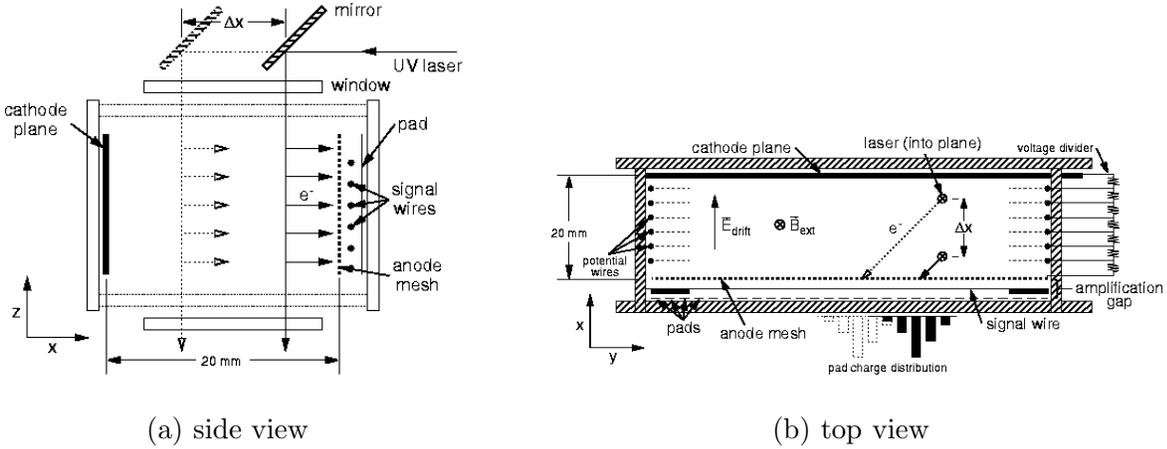

\centerline{
\subfigure[side view]{ \label{fig:side}
\epsfxsize=2.4in
\epsfbox{figs/sideview.epsi}}
\hspace{0.3in}
\subfigure[top view]{ \label{fig:top}
\epsfxsize=3.2in
\epsfbox{figs/topview.epsi}} }
\caption{Test drift chamber.  a) Side view: UV laser is displaced by a micrometer.  b) Top view: schematic pad charge distribution for magnetic deflection angle measurement.} \label{fig:chamber}
\end{figure}

To determine the principle features of the gases without distortion from the cylindrical geometry, a setup with uniform electric and magnetic fields is chosen (see \cite{bib:cal_setup} for a more complete description).  The setup consists of a drift chamber in the field of the former MIT cyclotron magnet.  A nitrogen laser produces ionization at $t = 0$ in the chamber.  The electrons drift under the influence of a uniform electric field through a wire mesh and into a proportional chamber as shown in Fig.~\ref{fig:side}.  Displacing the laser by $\Delta x$ with a mirror, the velocity parallel to the electric field $v_{||} = \frac{\Delta x}{\Delta t}$ is obtained.  The deflection by the magnetic field is calculated from the ``center of gravity'' of charge induced on pick-up strips behind the wires as seen in Fig.~\ref{fig:top}.
\vspace{.7in}

Differential measurements eliminate the influence of the amplification gap on the measured times.  The magnet can produce fields from $0 - 1.4$ T with $> 99 \%$ homogeneity over the chamber volume.  The chamber has a thick quartz window and can reliably hold pressures to 3 bar, as well as be pumped down to 0 bar absolute pressure before a measurement.  

\section{Measurements and scaling behavior for Ar dominated mixtures}
\begin{figure}[ht]
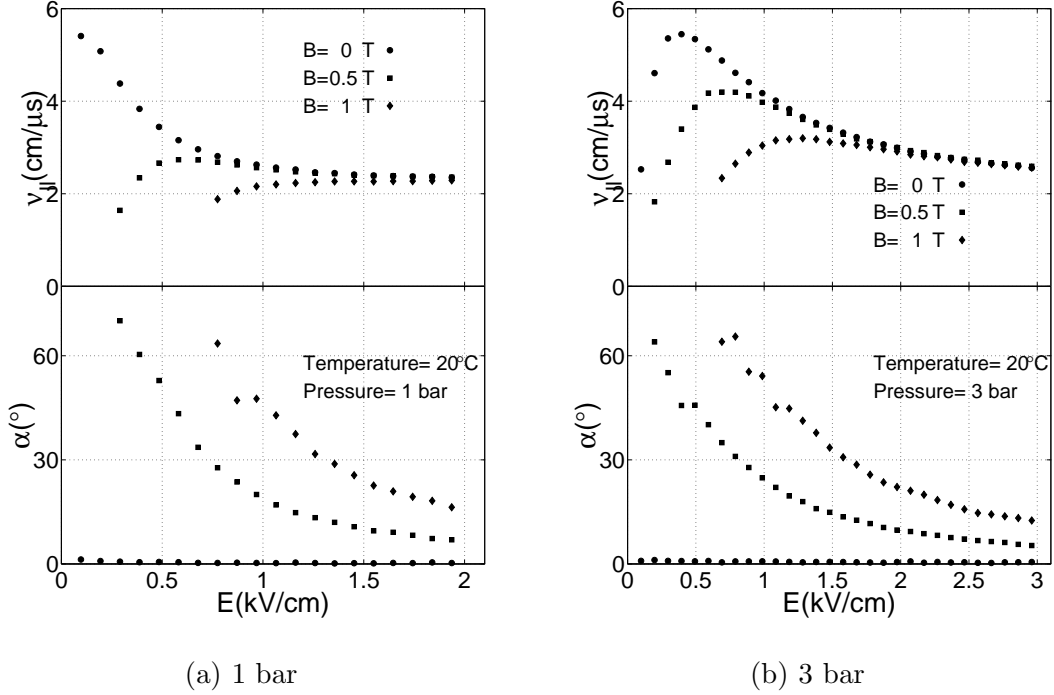

\centerline{
\subfigure[1 bar]{ \label{fig:p10_1bar}
\epsfxsize=2.5in
\epsfbox{figs/p10one.epsi}}
\hspace{0.3in}
\subfigure[3 bar]{ \label{fig:p10_3bar}
\epsfxsize=2.5in
\epsfbox{figs/p10three.epsi}} }
\caption{Measured drift velocities $v_{||}$ (parallel to $E$) and deflection angles $\alpha$ for various electric and magnetic field values.} \label{fig:p10_data}
\end{figure}

To establish regularities and simplify scaling laws, electron drift velocities and magnetic deflection angles were measured in a low admixture argon gas P10, Ar:CH$_4$ (90:10)\%, for different drift electric fields, magnetic fields, and pressures as shown in Figs.~\ref{fig:p10_data}.  

\begin{figure}[ht]
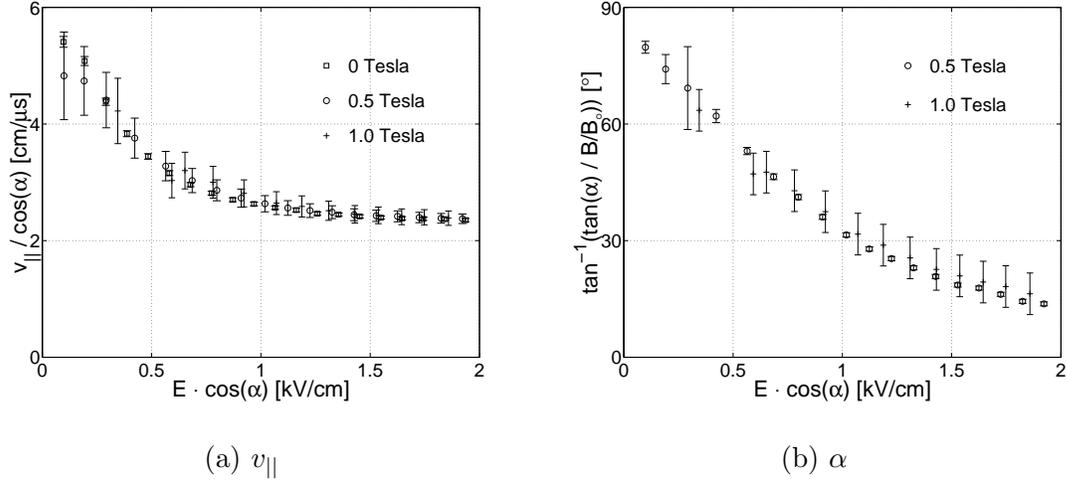

\centerline{
\subfigure[$v_{||}$]{ \label{fig:vbscale}
\epsfxsize=2.1in
\rotate{\epsfbox{figs/vbscale.epsi}}}
\hspace{0.3in}
\subfigure[$\alpha$]{ \label{fig:abscale}
\epsfxsize=2.1in
\rotate{\epsfbox{figs/abscale.epsi}}} }
\caption{Scaling for different magnetic fields in P10.  $B_\circ$ is 1 Tesla.} \label{fig:bscale}
\end{figure}

For relatively small magnetic field values, the total drift velocity $v_T$ will be approximately constant.  Hence, we expect
\begin{equation}
v_{||}(E\cos\alpha,B) = v_T(E,B=0),
\end{equation}
which is verified in Fig.~\ref{fig:vbscale}.  Similarly, because the Lorentz force is proportional to the magnetic field, the scaling for the magnetic deflection angles is expected to be
\begin{equation}
\frac{\tan\alpha_1}{B_1} = \frac{\tan\alpha_2}{B_2}
\end{equation}
as demonstrated in Fig.~\ref{fig:abscale}.

\begin{figure}[ht]
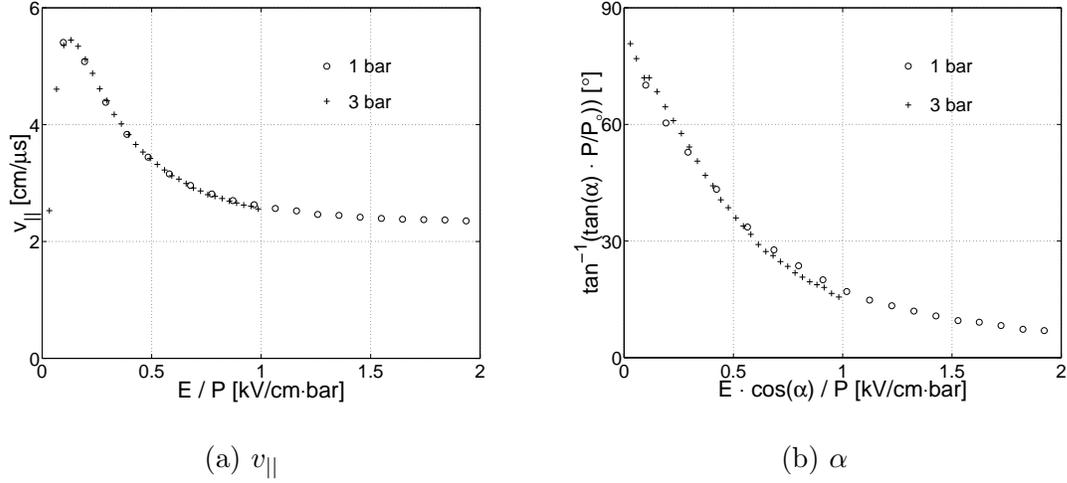

\centerline{
\subfigure[$v_{||}$]{ \label{fig:vpscale}
\epsfxsize=2.1in
\rotate{\epsfbox{figs/vpscale.epsi}}}
\hspace{0.3in}
\subfigure[$\alpha$]{ \label{fig:apscale}
\epsfxsize=2.1in
\rotate{\epsfbox{figs/apscale.epsi}}} }
\caption{Scaling for different pressures in P10.  $P_\circ$ is 1 bar.} \label{fig:pscale}
\end{figure}

For different pressures $P$, one expects a behavior governed by the mean free path of electrons in the gas, $\lambda \propto \frac{1}{P}$, and therefore scaling with $\frac{E}{P}$ as verified in Figs.~\ref{fig:pscale}.

\section{Search for a quasi-linear gas for a cylindrical geometry}
\begin{figure}[ht]
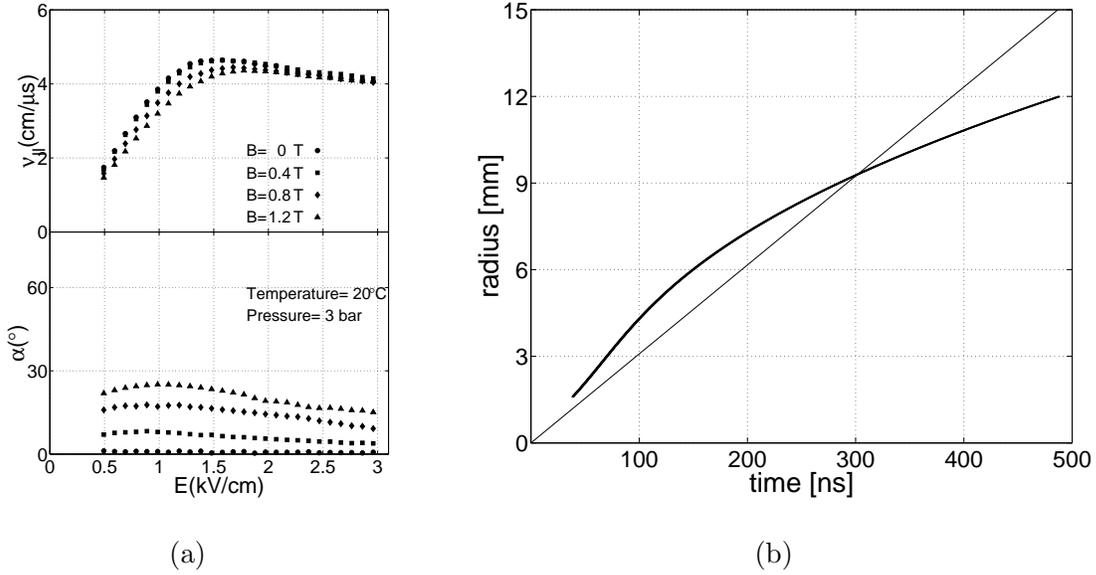

\centerline{
\subfigure[]{ \label{fig:arco2_data}
\epsfxsize=2.0in
\epsfbox{figs/93_7.epsi}}
\hspace{0.3in}
\subfigure[]{ \label{fig:arco2_rt}
\epsfxsize=2.6in
\rotate{\epsfbox{figs/93_7_rt.epsi}}} }
\caption{(a) Velocities and angles for Ar:CO$_2$ (93:7)\%.  (b) Corresponding space-time relation in a cylindrical geometry.} \label{fig:arco2}
\end{figure}

Fig.~\ref{fig:arco2_data} displays the drift features of Ar:CO$_2$ (93:7)\% measured in our setup at 3 bar.  The non-linearity of the corresponding space-time relation in Fig.~\ref{fig:arco2_rt} is a result of the change of drift velocity close to the drift tube wall, in the $0.5 - 1$ kV/cm range.  These low velocities are responsible for long times at large radius.

\begin{figure}[ht]
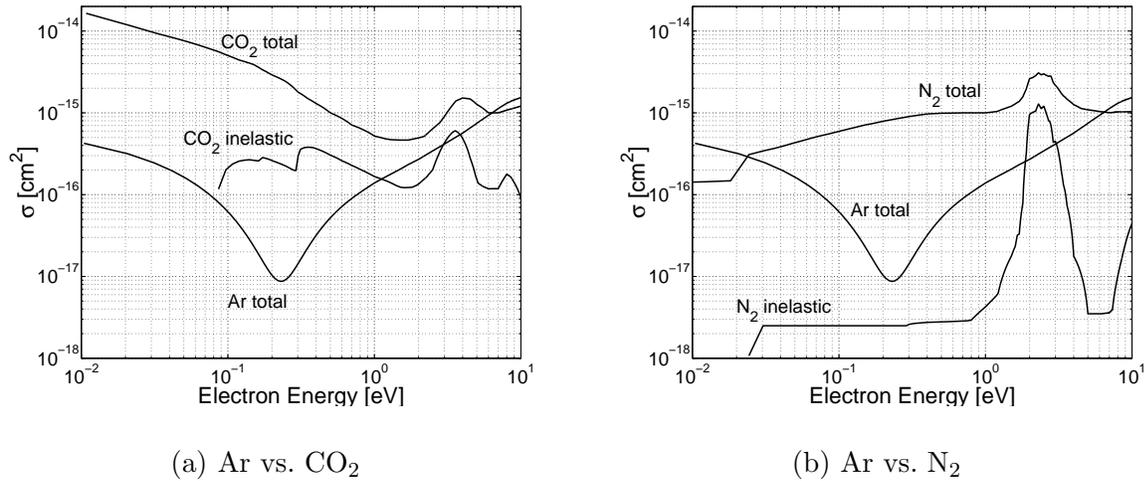

\centerline{
\subfigure[Ar vs. CO$_2$]{ \label{fig:co2_xsec}
\epsfxsize=2.1in
\rotate{\epsfbox{figs/co2_xsec.epsi}}}
\hspace{0.3in}
\subfigure[Ar vs. N$_2$]{ \label{fig:n2_xsec}
\epsfxsize=2.1in
\rotate{\epsfbox{figs/n2_xsec.epsi}}} }
\caption{Energy dependent electron-gas interaction cross-sections for argon compared with carbon dioxide and nitrogen\cite{bib:magboltz}.} \label{fig:xsecs}
\end{figure}

The improvement of the Ar:CO$_2$ (93:7)\% mixture is based on considering the electron-gas interaction cross-sections of admixture gases (Figs.~\ref{fig:xsecs}).  Due to quantum mechanical effects, these cross-sections are a function of energy.  For example, the cross-section of argon has a minimum at about 0.23 eV, meaning that when electrons have this energy, they interact less strongly with the Ar atoms.  This explains the maximum in drift velocity of gas mixtures containing argon (Figs.~\ref{fig:p10_data}).  Carbon dioxide has a large inelastic cross-section at low energy which tends to keep the mean electron energy low and around that of the argon cross-section minimum \cite{bib:inelastic}.  This strongly increases non-linearity, but the smaller average time between collisions leads to smaller angles.  

To obtain a more constant electron velocity at low electric field, the amount of carbon dioxide was reduced.  The addition of nitrogen to a mixture of argon and carbon dioxide was advantageous at higher electric fields because nitrogen has a large elastic cross-section.  This helps to diminish the effect of the minimum in the argon cross-section (Fig.~\ref{fig:n2_xsec}).  Thus, the addition of nitrogen should flatten the drift velocity curve and straighten the space-time relation.  The negative effect is a modest increase in deflection angle.  

\section{Results from new mixtures}
\begin{figure}[ht]
\centerline{
\subfigure[Ar:N$_2$:CO$_2$ (94:2:4)\%]{ \label{fig:94_2_4_dat}
\epsfxsize=2.5in
\epsfbox{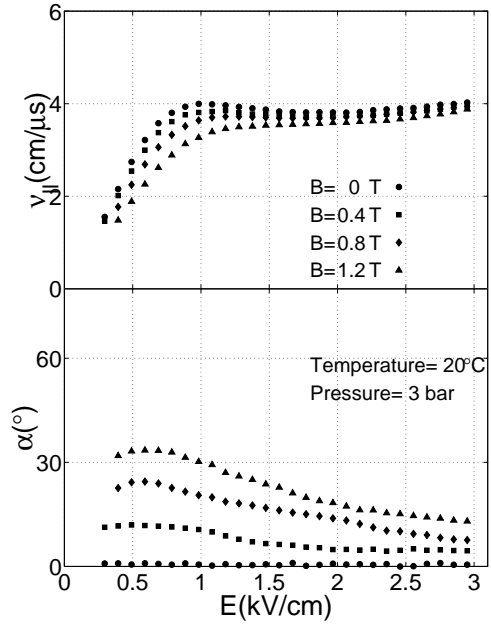}}
\hspace{0.3in}
\subfigure[Ar:N$_2$:CO$_2$ (92:6:2)\%]{ \label{fig:92_6_2_dat}
\epsfxsize=2.5in
\epsfbox{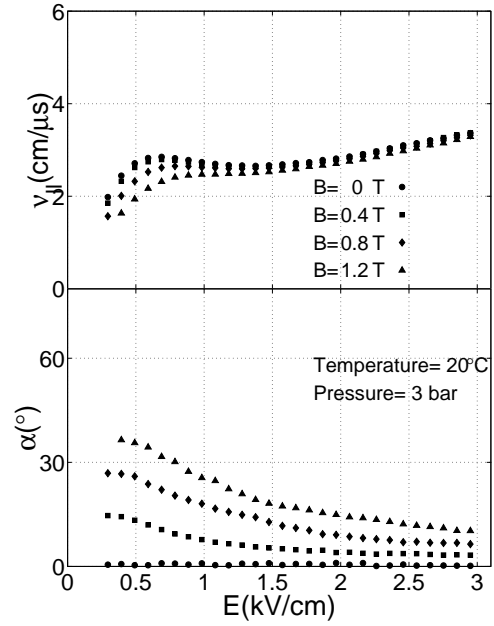}} } 
\centerline{
\subfigure[Ar:N$_2$:CO$_2$ (94:2:4)\%]{ \label{fig:94_2_4_rt}
\epsfxsize=2.1in
\rotate{\epsfbox{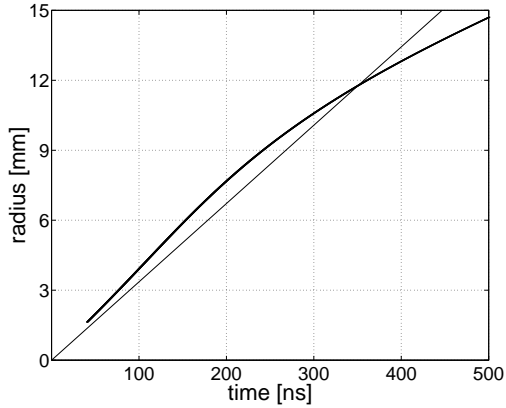}}}
\hspace{0.3in}
\subfigure[Ar:N$_2$:CO$_2$ (92:6:2)\%]{ \label{fig:92_6_2_rt}
\epsfxsize=2.1in
\rotate{\epsfbox{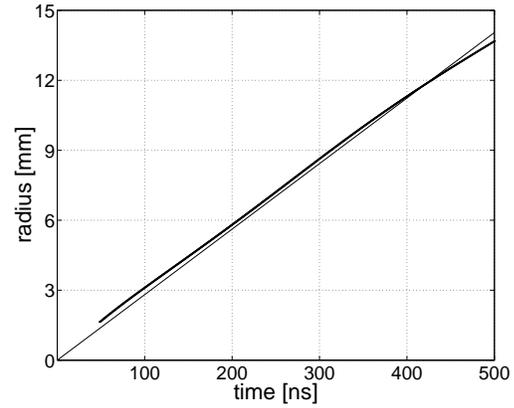}}} }
\caption{a,b) Measured drift velocities and deflection angles.  c,d) Corresponding space-time relation in a cylindrical geometry.} \label{fig:goodgases}
\end{figure}

The data in Figs.~\ref{fig:goodgases} demonstrate the predicted features from the discussion above.  Figs.~\ref{fig:94_2_4_dat}~and~\ref{fig:92_6_2_dat} show mixtures which do not contain any hydrocarbon admixture.  The magnetic deflection angles $\alpha$ are slightly higher than the angles in Fig.~\ref{fig:arco2_data} but the velocities lead to much better space-time curves as demonstrated in Figs.~\ref{fig:94_2_4_rt}~and~\ref{fig:92_6_2_rt} by adapting the measured data to a cylindrical geometry.  

\section{Discussion of accuracy}
\begin{figure}[ht]
\centerline{
\subfigure[two layers]{ \label{fig:2tubes}
\epsfxsize=2.5in
\epsfbox{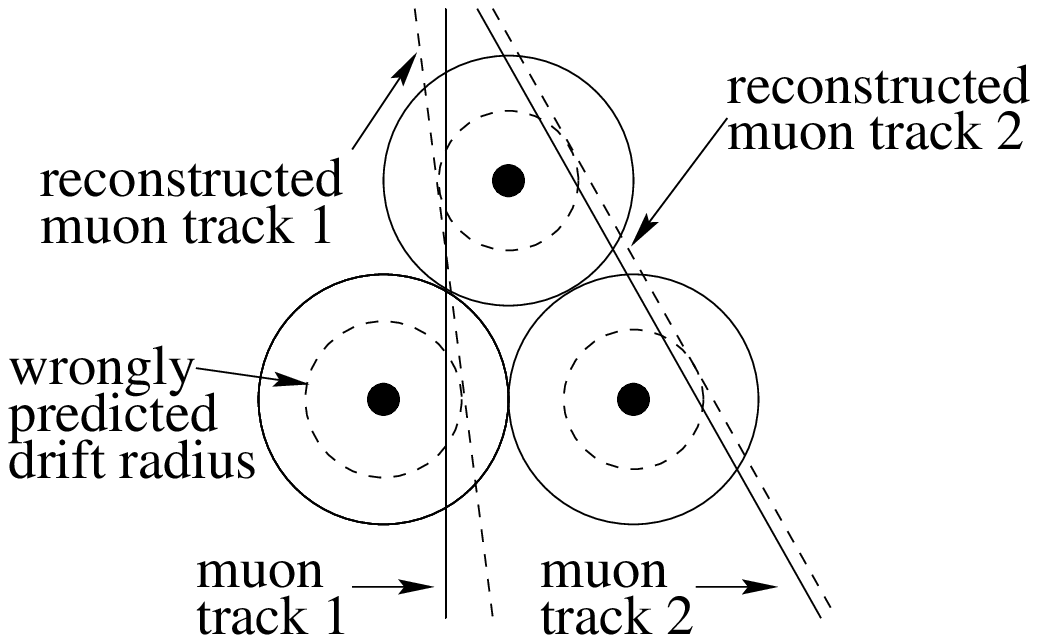}}
\hspace{0.3in}
\subfigure[three layers]{ \label{fig:3tubes}
\epsfxsize=2.5in
\epsfbox{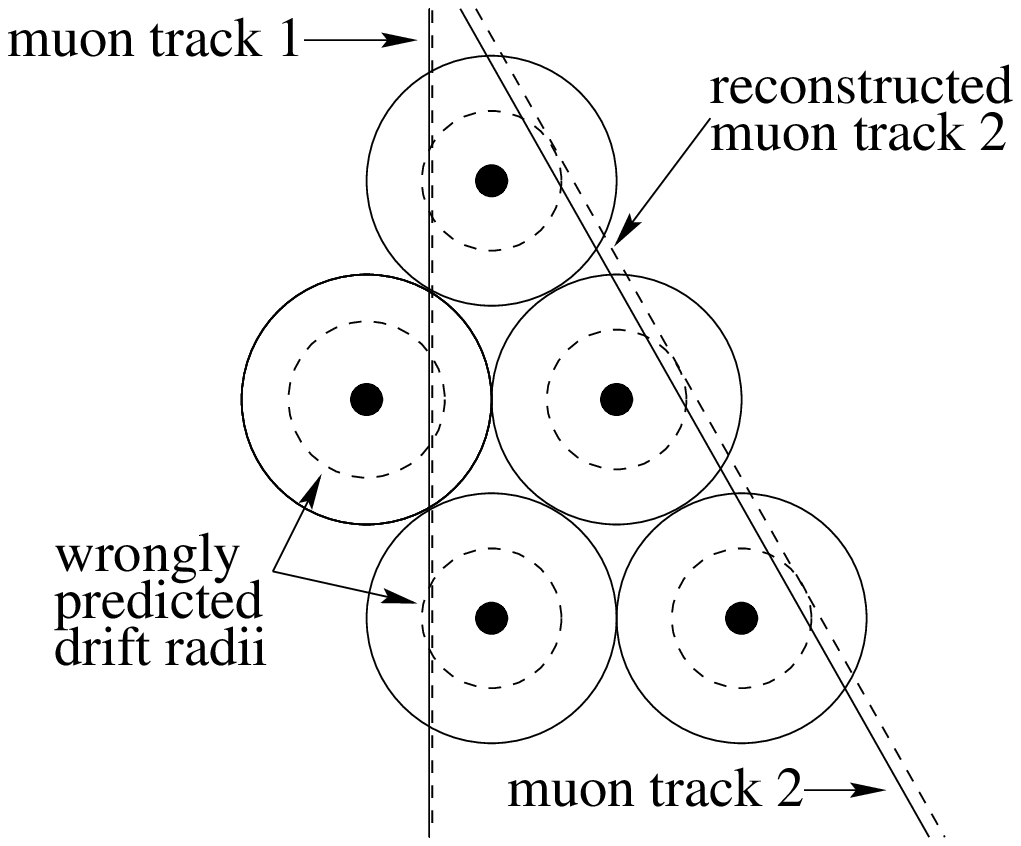}} }
\caption{Schematic describing resolution degradation due to non-linearities in the space-time curve.} \label{fig:tubes}
\end{figure}

\begin{figure}[ht]
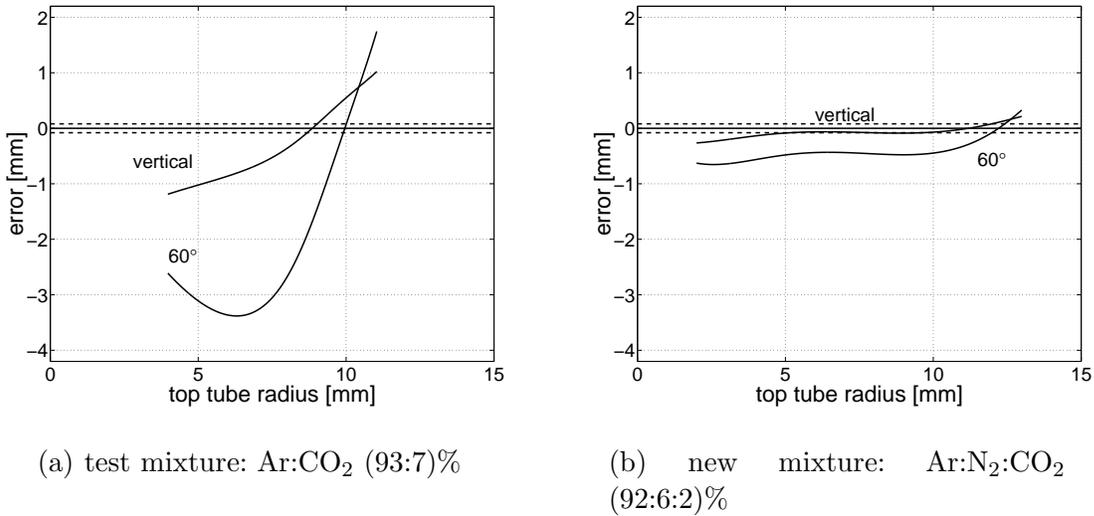

\centerline{
\subfigure[test mixture: Ar:CO$_2$ (93:7)\%]{ \label{fig:93_7_err}
\epsfxsize=2.1in
\rotate{\epsfbox{figs/err_93_7.epsi}}}
\hspace{0.3in}
\subfigure[new mixture: Ar:N$_2$:CO$_2$ (92:6:2)\%]{ \label{fig:92_6_2_err}
\epsfxsize=2.1in
\rotate{\epsfbox{figs/err_92_6_2.epsi}}} }
\caption{Reconstruction errors in the used gas and the proposed gas.  The dashed line represents the ATLAS resolution goal of 80 $\mu$m.} \label{fig:errors}
\end{figure}

To see the benefits of a linear drift gas, a simple model consisting of a two-layer set of drift tubes is considered (Fig.~\ref{fig:2tubes}).  Muon track 1 is the actual path of the muon, but due to non-linearities in the space-time relation of the gas, this track will be incorrectly reconstructed.  At 60$^\circ$ from vertical, muon track 2 is also affected by these non-linearities.  Fig.~\ref{fig:3tubes} shows the systematic errors when three layers of drift tubes are used.  The third layer helps to pull the reconstructed muon track 1 back to vertical, but the reconstructed muon track 2 continues to suffer from the incorrectly predicted radii.  Using the space-time curves from Figs.~\ref{fig:arco2_rt},~\ref{fig:94_2_4_rt},~and~\ref{fig:92_6_2_rt}, horizontal deviations in the muon track reconstruction were calculated for muons impinging vertically and at 60$^\circ$ from vertical.  These errors are shown in Figs.~\ref{fig:errors}.  It is clear that the reconstruction error in a gas is strongly related to the linearity of its space-time relation.

\section{Conclusions}
We show that simple relations can be used to scale mean drift velocity and magnetic deflection angle data for magnetic fields $<$ 1.2 T and pressures $<$ 3 bar.  Transforming measured data in homogenous fields to a cylindrical geometry lead to a gas mixture with a nearly linear space-time relation.

With knowledge of gas admixtures, definite improvements over the present ATLAS MDT test mixture Ar:CO$_2$ (93:7)\% were found.  Ar:N$_2$:CO$_2$ (94:2:4)\% gives roughly a factor three improvement and Ar:N$_2$:CO$_2$ (92:6:2)\% has an order of magnitude smaller corrections and may provide a chance to improve accuracy in the large ATLAS detector.

\ack{We wish to acknowledge Professors G. Herten, S. Ahlen, and F. Taylor for their contributions to this project.  We thank Prof S. Biagi for the recent version of Magboltz.  We acknowledge the Laboratory for Nuclear Science for support and M. Grossman and F. Cot\'e for technical assistance.  The work was supported under DOE contract \#DE-FC02-94 ER 40818.}

\end{document}